\documentclass{aa}             

\input epsf

\begin{document}

\def\EW{EW\,Ori}    
\def\kms{\ifmmode{\rm km\thinspace s^{-1}}\else km\thinspace s$^{-1}$\fi}
\def\cm{\ifmmode{\rm cm\thinspace^{-1}}\else cm\thinspace$^{-1}$\fi}
\def\Msun{$M_{\sun}$}
\def\Rsun{R_{sun}}
\def\Lsun{L_{sun}}
\def\feh{$[\mathrm{Fe/H}]$}
\def\meh{$[\mathrm{M/H}]$}
\def\afe{$[\mathrm{\alpha/Fe}]$}
\def\ione{\,{\sc i}}
\def\itwo{\,{\sc ii}}

\title{
Absolute dimensions of solar-type eclipsing binaries. III.
\thanks{Based on observations carried out at the Str{\"o}mgren Automatic 
Telescope (SAT) and the 1.5m telescope (62.L-0284) at ESO, La Silla
}}
\subtitle{EW\,Orionis: \\Stellar evolutionary models tested by a G0~V system.
\thanks{
Table~11 is available in electronic form
at the CDS via anonymous ftp to cdsarc.u-strasbg.fr (130.79.128.5)
or via http://cdsweb.u-strasbg.fr/cgi-bin/qcat?J/A+A/}
}
\author{
J.V. Clausen     \inst{1}
\and
H. Bruntt        \inst{2,3}
\and
E.H. Olsen       \inst{1}
\and
B.E. Helt        \inst{1}
\and
A. Claret        \inst{4}
}
\offprints{J.V.~Clausen, \\ e-mail: jvc@nbi.ku.dk}

\institute{
Niels Bohr Institute, Copenhagen University,
Juliane Maries Vej 30,
DK-2100 Copenhagen {\O}, Denmark
\and
Sydney Institute for Astronomy, School of Physics, University of Sydney,
NSW 2006, Australia
\and
Observatoire de Paris, LESIA, 5 Place Jules Janssen,
95195 Meudon, France
\and
Instituto de Astrof\'isica de Andaluc\'ia, CSIC,
Apartado 3004, E-18080 Granada, Spain
}

\date{Received 20 November 2009 / Accepted 14 December 2009} 

\titlerunning{EW\,Ori}
\authorrunning{J.V. Clausen et al.}

\abstract
{Recent studies of inactive and active solar-type binaries suggest
that chromospheric activity, and its effect on envelope convection,
is likely to cause significant radius and temperature discrepancies.
Accurate mass, radius, and abundance determinations from additional
solar-type binaries exhibiting various levels of activity are needed 
for a better insight into the structure and evolution of these stars.}
{We aim to determine absolute dimensions and abundances for the
G0~V detached eclipsing binary \EW, and to perform a detailed
comparison with results from recent stellar evo\-lu\-tio\-nary models.}
{$uvby$ light curves and $uvby\beta$ standard photometry were obtained with
the Str\"omgren Automatic Telescope, published radial velocity observations 
from the CORAVEL spectrometer were reanalysed, and high-resolution spectra 
were observed at the FEROS spectrograph; all are/were ESO, La Silla facilities.
State-of-the-art methods were applied for the photometric and spectroscopic
analyses.}
{Masses and radii that are precise to 0.9\% and 0.5\%, respectively, have
been established for both components of \EW.
The 1.12 \Msun\ secondary component reveals weak Ca\,\itwo\ H and K emission 
and is probably mildly active; no signs of activity are seen for the 
1.17 \Msun\ primary.
We derive an \feh\ abundance of $+0.05 \pm 0.09$ and similar abundances for
Si, Ca, Sc, Ti, Cr, and Ni.
Yonsai-Yale and Granada solar-scaled evolutionary models for the observed 
metal abundance reproduce the components fairly well at an age of 
$\approx$2 Gyr.
Perfect agreement is, however, obtained at an age of 2.3 Gyr for a combination 
of $a)$ a slight downwards adjustment of the envelope mixing length parameter
for the secondary, as seen for other active solar-type stars, and
$b)$ a slightly lower helium content than prescribed by the $Y-Z$ relations
adopted for the standard model grids.  
The orbit is eccentric ($e = 0.0758 \pm 0.0020$), and apsidal motion with a
62\% relativistic contribution has been detected. The apsidal motion period is
$U = 16\,300 \pm 3\,900$ yr, and the inferred mean central density concentration
coefficient, log($k_2$) = $-1.66 \pm 0.30$, agrees marginally with
model predictions.
The measured rotational velocities, $9.0 \pm 0.7$ (primary) and
$8.8 \pm 0.6$ (secondary) \kms, are in agreement with both the synchronous
velocities and the theoretically predicted pseudo-synchronous velocities.
Finally, the distance ($175 \pm 7$ pc), age, and center-of mass velocity 
(6 \kms) exclude suggested membership of the open cluster Collinder 70.
}
{
\EW\ now belongs to the small group of solar-type eclipsing binaries with
well-established astrophysical properties. 
}
\keywords{
Stars: evolution --
Stars: fundamental parameters --
Stars: abundances --
Stars: binaries: eclipsing --
Techniques: photometric --
Techniques: spectroscopic}

\maketitle

\section{Introduction}
\label{sec:intro}

\begin{figure*}
\epsfxsize=185mm
\epsfbox{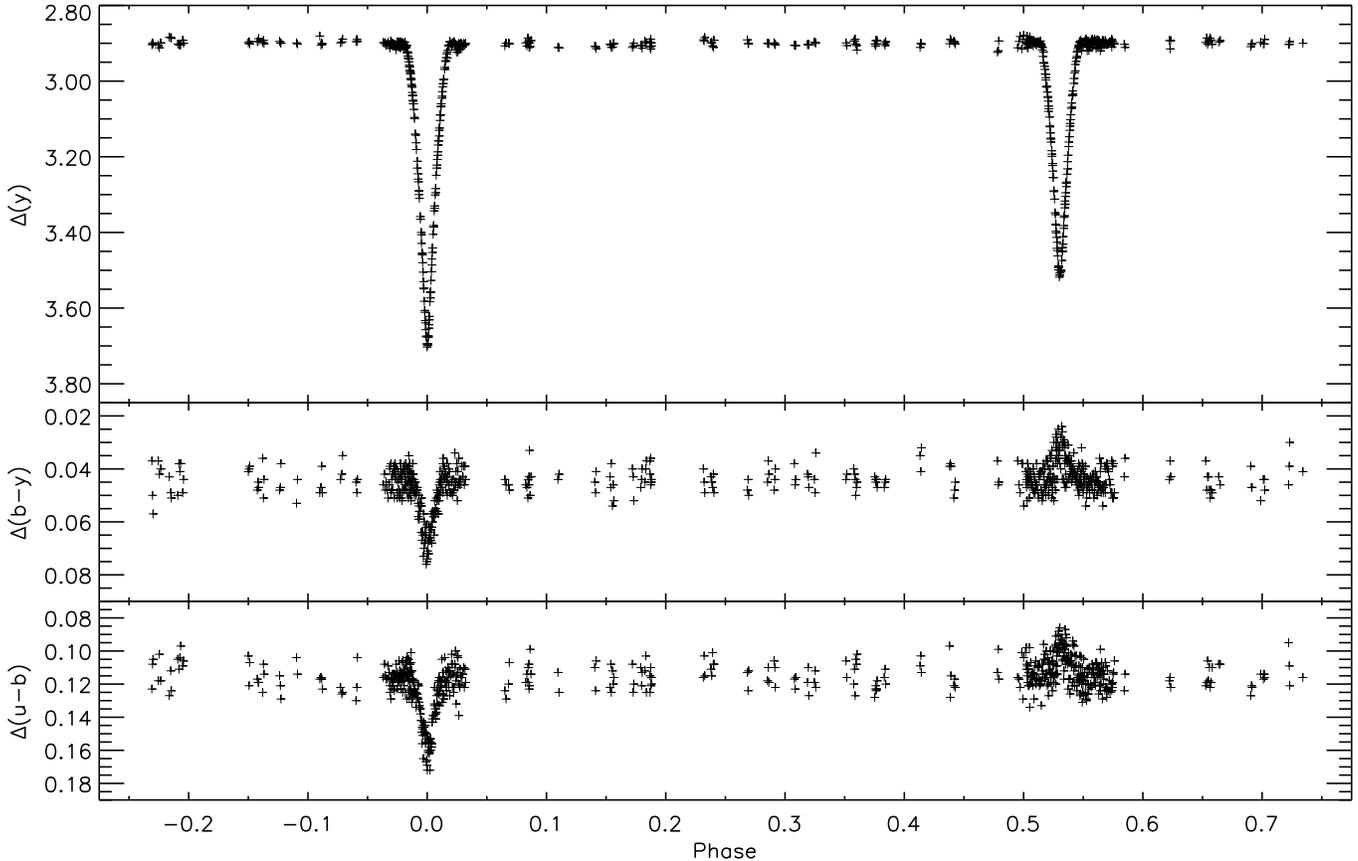}
\caption[]{\label{fig:ewori_lc}
$y$ light curve and $b-y$ and $u-b$ colour curves (instrumental system)
for EW\,Ori.
}
\end{figure*}

Recent studies of inactive and active solar-type binaries suggest that
chromospheric activity, and its effect on envelope convection, is likely
to cause significant radius and temperature discrepancies,
which can be removed by adjusting the model mixing length parameter 
downwards (Clausen et al. \cite{cbc09}, hereafter CBC09, and references herein). 
In a study of the F8/G0~V eclipsing binary \object{V636\,Cen}, CBC09
compared the properties of 11 solar-type 
binaries (at least one component in the 0.80--1.10 $M_{\sun}$ mass interval)
with well-established dimensions, which -- with one exception --
fall in two groups:
Four long period, slowly rotating inactive systems,
which seem to be fitted reasonably well by solar-scaled models.
And six systems, which exhibit intrinsic variation, have short orbital
periods and high rotational velocities, and which can not be reproduced
by solar-scaled models. In addition, the authors found that the sample
lends support to theoretical 2D radiation hydrodynamics studies, which predict 
a slight decrease of the mixing length parameter with increasing 
temperature/mass for $\it inactive$ main sequence stars.
More binaries are, however, needed for a calibration in terms
of physical parameters and level of activity.

We are presently undertaking analyses of several eclipsing binaries with 
solar-type components; see CBC09. In order to make critical tests of
stellar evolutionary models, abundance determinations are in general
included. In this paper we present results
for the well-known system \EW\, based on new observations.

\begin{table*}
\caption[]{\label{tab:ewori_std}
Photometric data for \EW\ and the comparison stars.
}
\begin{minipage}{\textwidth}
\begin{center}
\begin{tabular}{lllrrrrrrrrrrrr} \hline
\hline\noalign{\smallskip}
Object&Sp. Type&Ref.     &$V$&$\sigma$&$b-y$&$\sigma$&$m_1$&$\sigma$&$c_1$ &$\sigma$&N($uvby$)&$\beta$&$\sigma$&N($\beta$)\\
\noalign{\smallskip}
\hline
\noalign{\smallskip}
 EW\,Ori  & G0~V$^{\mathrm{a}}$       & C10    & 9.902  & 7& 0.390 & 5& 0.186 & 8& 0.364 &10&324 & 2.613   & 8 &25\\    
          &            & C10    &10.510  & 7& 0.379 & 5& 0.181 & 4& 0.369 & 4&  4 &         &   &  \\ 
          &            & HH75   & 9.905  & 7& 0.387 & 5& 0.201 & 1& 0.350 &10&  2 &         &   &  \\
          &            & L02    & 9.924  &  & 0.407 &  & 0.175 &  & 0.353 &  &  1 & 2.629   &   & 1\\
\hline
 HD34658  & F3~III/IV$^{\mathrm{b}}$  & C10    & 5.334  & 5& 0.258 & 3& 0.196 & 5& 0.668 & 6&180 & 2.691   & 4 &15\\ 
          &            & SP65   &        &  & 0.268 &  & 0.177 &  & 0.682 &  &  2 &         &   &  \\
          &            & C66    &        &  &       &  &       &  &       &  &    & 2.685   &   & 7\\
          &            & GO7677 & 5.334  & 2& 0.260 & 2& 0.189 & 3& 0.671 & 5&  4 & 2.685   & 6 & 3\\
          &            & AF90   & 5.326  &12& 0.263 &20& 0.184 &18& 0.659 &23&  2 & 2.676   & 3 & 2\\
\hline
 HD34745  & F7~V$^{\mathrm{b}}$       & C10    & 6.999  & 6& 0.343 & 4& 0.171 & 7& 0.363 & 8&236 & 2.618   & 8 &17\\
          &            & O83    & 6.998  & 6& 0.351 & 1& 0.168 & 5& 0.360 & 1&  2 &         &   &  \\
          &            & O94    & 7.008  & 4& 0.350 & 3& 0.157 & 4& 0.367 & 5&  1 & 2.614   & 6 & 1\\  
\hline
 HD35638  & F5~V$^{\mathrm{b}}$       & C10    & 7.645  & 6& 0.290 & 4& 0.164 & 7& 0.420 & 8&172 & 2.651   & 8 &16\\
          &            & O83    & 7.648  & 5& 0.297 & 3& 0.158 & 2& 0.417 & 1&  2 &         &   &  \\
          &            & O94    & 7.653  & 5& 0.295 & 3& 0.151 & 4& 0.430 & 6&  1 & 2.648   & 6 & 1\\ 
\hline
\end{tabular}
\begin{list}{}{}
\item[$^{\mathrm{a}}$] Popper et al. (\cite{plft86}); spectral type of the primary component.
\item[$^{\mathrm{b}}$] Houk \& Swift (\cite{houk99}).
\end{list}
\end{center}
\textsc{NOTE 1:}
References are:
AF90 = Arellano Ferro et al. (\cite{af90}),
C10 = This paper,
C66 = Crawford et al. (\cite{c66}),
GO7677 = Gr{\o}nbech \& Olsen (\cite{go76},\cite{go77}) (revised catalogue),
HH75   = Hilditch \& Hill (\cite{hh75}),
L02    = Lacy (\cite{l02}),
O83  = Olsen (\cite{olsen83}),   
O94  = $uvby$: Olsen (\cite{olsen94}), $\beta$: Olsen (unpublished),   
SP65 = Str\"omgren \& Perry (\cite{sp2}).

\textsc{NOTE 2:}
For EW\,Ori, the $uvby\beta$ information by C10 (first line), HH75, and L02 is 
the mean value outside eclipses. For C10 (second line) indices during the total
part of secondary eclipse are included.

\textsc{NOTE 3:}
N is the total number of observations used to form the mean values, and
$\sigma$ is the rms error (per observation) in mmag.
\end{minipage}
\end{table*}

\section{EW\,Ori}
\label{sec:ewori}

\object{EW\,Ori} (HD~287727, $m_{V}$ = 9.90, Sp. type G0V, $P$ = 6\fd94),
is a well detached, double-lined eclipsing binary with 1.17 and 1.12 \Msun\ 
main-sequence components in a slightly eccentric ($e$ = 0.0758) orbit.
The eclipsing nature of \EW\ was discovered by Hoffmeister (\cite{h30}), and
Lause (\cite{l37}) established the first ephemeris.
Much later, \EW\ was found to be double-lined (Lacy \cite{l84}).
Several times of minima, back to 1937, have been published, as well as
photometric indices. 
Popper et al. (\cite{plft86}) obtained photoelectric
($V,R$) light curves and spectrographic material, and they determined absolute
dimensions. 
Improved spectroscopic elements, based on CORAVEL radial velocities, were
later published by Imbert (\cite{imbert02}).

\EW\ has been included in samples of eclipsing binaries used for tests of
stellar evolutionary models (Pols et al. \cite{pols97}, 
Lastennet \& Valls-Gabaud \cite{lastennet02}), 
determination of the helium-to-metal enrichment ratio 
(Ribas et al. \cite{ribas00}), and as a (possible) test of general relativity 
through apsidal motion (Gim\'enez \cite{agc85}, 
Wolf et al. \cite{wolf97,wolf09}). 
It is listed as a possible member of the open cluster \object{Collinder~70} by 
Sahade \& Davila (\cite{sd63}) and Gim\'enez \& Clausen (\cite{gc96}).

Although \EW\ is already well-studied, there is still room for significant
improvements.
In this paper we present analyses of new $uvby$ light curves, leading to
much more accurate radii, we derive chemical abundances from
high-resolution spectra, and we perform a detailed comparison with current
stellar evolutionary models.
We refer to the more massive, larger component as the 
primary ($p$) component, which for the ephemeris we adopt 
(Eq.~\ref{eq:ewori_eph}) is eclipsed at phase 0.0.

\section{Photometry}
\label{sec:phot}

Below, we present the new photometric material for \EW\ and refer to Clausen 
et al. (\cite{jvcetal01}; hereafter CHO01) for further details on observation 
and reduction procedures, and determination of times of minima.

\subsection{Light curves}
\label{sec:lc}

The differential $uvby$ light curves of \EW\ were observed at the
Str{\"o}mgren Automatic Telescope (SAT) at ESO, La Silla
and its 6-channel $uvby\beta$ photometer on 70 nights between
November 2000 and December 2002 (JD2451854--2452609). 
They contain 624 points per band with most phases covered at least twice.
The observations were done through an 18 arcsec diameter circular
diaphragm at airmasses between 1.2 and 2.0. 
\object{HD~34658} (HR~1746, NSV~1922), \object{HD~34745}, 
and \object{HD~35638} --
all within a few degrees from \EW\ on the sky -- were used as 
comparison stars and were all found to be constant within a few mmag; 
see Table~\ref{tab:ewori_std}. 
The light curves are calculated relative to HD~34745, but all comparison 
star observations were used, shifting them first to the same light level.
The average accuracy per point is about 6 mmag ($ybv$) and 7 mmag ($u$).

As seen from Fig.~\ref{fig:ewori_lc}, \EW\ is well detached with 
$y$ eclipse depths of about 0.8 and 0.6 mag, respectively. 
The secondary eclipse is total (duration of totality about 10 minutes) 
and occurs at phase 0.5305.
The light curves (Table 11) will only be available in electronic form.

\subsection{Standard photometry}
\label{sec:std}

Standard $uvby\beta$ indices for \EW\ and the three comparison stars,
observed and derived as described by CHO01, are presented in 
Table~\ref{tab:ewori_std}. As seen, the indices are based on many
observations and their precision is high.
For comparison, we have included published photometry from 
other sources.
In general, the agreement is good, but individual differences larger 
than the quoted errors occur; we have used the new results 
for the analysis of \EW.

\subsection{Times of minima, ephemeris, and apsidal motion}
\label{sec:tmin}

\begin{table}
\caption[]{\label{tab:ewori_tmin}Times of primary (P) and secondry (S)
minima of EW\,Ori determined from the $uvby$ observations.
}
\begin{minipage}{\columnwidth}
\centering
\renewcommand{\footnoterule}{}  
\begin{tabular}{llcr} \hline
\hline\noalign{\smallskip}
HJD           & rms    & Type &   O-C(P)\footnote{Calculated for the ephemeris given in Eq.~\ref{eq:ewori_eph}}\\
$-$ 2\,400\,000 &      &      &   Phase(S) \\
\noalign{\smallskip}
\hline
\noalign{\smallskip}
51877.80164   &0.00007 & P    &   0.00022  \\
51884.73864   &0.00018 & P    &   0.00038  \\
51898.61176   &0.00009 & P    & $-0.00019$ \\
51860.67097   &0.00034 & S    &   0.53051  \\
51978.59500   &0.00100 & S    &   0.53018  \\
51985.53400   &0.00020 & S    &   0.53049  \\
52304.62875   &0.00011 & S    &   0.53049  \\
\hline
\end{tabular}
\end{minipage}
\end{table}

Three times of the primary minimum and four of the secondary have been 
determined
from the $uvby$ light curve observations; see Table~\ref{tab:ewori_tmin}.
A complete list of earlier times of minima was kindly provided by
Kreiner and has been included in the ephemeris and apsidal motion analyses;
see Kreiner et al. (\cite{kreiner01}) and Kreiner 
(\cite{kreiner04})\footnote{{\scriptsize\tt http://www.as.ap.krakow.pl/ephem}}.

Separate weighted linear least squares fits to the times of primary and
secondary minima lead to slightly different orbital periods of
$6\fd93684324 \pm 0.00000044$ and 
$6\fd93684514 \pm 0.00000018$, respectively, and nearly identical
results are obtained if only photoelectric and CCD data are used.
Whereas this difference indicates a slow apsidal motion, we have
adopted the result from the primary minima for the analyses of the
$uvby$ light curves and radial velocities in this paper:

\begin{equation}
\label{eq:ewori_eph}
\begin{tabular}{r r c r r}
{\rm Min \, I} =  & 2451877.80142 & + & $6\fd 93684324$ &$\times\; E$ \\
                  &      $\pm   9$&   &    $\pm     44$ &             \\
\end{tabular}
\end{equation}

\noindent
Within uncertainties, the same ephemeris is obtained from the
analyses of the $uvby$ light curves (Sect.~\ref{sec:phel}) if 
the period and the epoch are included as adjustable parameters.

From a weighted least squares analysis, following the formalism by
Gim\'enez \& Garcia-Pelayo (\cite{ggp83}) and Gim\'enez \& Bastero
(\cite{gb95}), we obtain the apsidal motion parameters presented
in Table~\ref{tab:ewori_aps}. The orbital inclination $i$ and
eccentricity $e$ were fixed at the values derived from the photometric
analysis (Table~\ref{tab:ewori_phel}). 
As seen, a slow but sig\-ni\-fi\-cant motion has been detected, but the apsidal 
motion period is still very uncertain. Our results agree with those published
by Wolf et al. (\cite{wolf97});
recently, Wolf et al. (\cite{wolf09}) obtained 
$\omega_1 = 0.00057 \pm 0.00004$ \degr/cycle.
We note that the sidereal period is equal to the mean of the periods determined
above from primary and secondary eclipses, respectively, and that 
the longitude of periastron, $\omega$, derived for the epoch of the 
$uvby$ observations is identical to the
result obtained from the light curve analyses (Sect.~\ref{sec:phel}).
The contribution from general relativity (Gim\'enez \cite{agc85}, Eqs.~3,4)
is 0.00026 \degr/cycle, or about 62\% of the observed rate. 
Within the rather large uncertainties, the derived mean central density 
concentration coefficient\footnote {see e.g. Gim\'enez (\cite{agc85}) for the
definition of $k_2$ and for references to the 'classical' papers}
 log($k_2$) = $-1.66 \pm 0.30$ marginally agrees 
with predictions from evolutionary models, -1.91 
(Table~\ref{tab:ewori_ac}, model set \#3).

\begin{table}
\caption[]{\label{tab:ewori_aps}
Apsidal motion parameters for EW\,Ori derived from all
available times of minima.
}
\begin{center}
\begin{tabular}{ll} \hline
\hline\noalign{\smallskip}
Parameter    & Value and rms error                    \\
\noalign{\smallskip}
\hline
\noalign{\smallskip}
$i$ (\degr)                & 89.86 (assumed)             \\ 
$e$                        & 0.0758 (assumed)            \\ 
$T_0$                      & $2427543.46225 \pm 0.00079 $  \\ 
$P_{anomalistic}$ (d)      & $6.9368523 \pm 0.0000018$    \\ 
$P_{sidereal}$    (d)      &  6.9368442                  \\ 
$\omega_0$ (\degr)         & $307.6 \pm 0.4$           \\  
$\omega_1$ (\degr / cycle) & $0.00042 \pm 0.00010$       \\
$U$ (yr)                   & $16300 \pm 3900 $           \\
\hline
\end{tabular}
\end{center}
\end{table}

\section{Photometric elements}
\label{sec:phel}

\begin{table}
\caption[]{\label{tab:ewori_jktebop_vh}
Photometric solutions for EW\,Ori from the JKTEBOP code.}
\begin{center}
\begin{tabular}{lrrrr} \hline
\hline\noalign{\smallskip}
                     &     $y$    &       $b$  &       $v$  &   $u$\\                   
\noalign{\smallskip}
\hline
\noalign{\smallskip}
$i$ \, (\degr)       &  89.89     &   89.84    &   89.87    &   89.85\vspace{-0.8mm}\\   
                     & $\pm 5$    &  $\pm 4$   &  $\pm 5$   &  $\pm 5$\\                 

$e\cos \omega$       &  0.04778   &   0.04775  &   0.04774  &   0.04767\vspace{-0.8mm}\\ 
                     &$\pm    3$  & $\pm    3$ & $\pm    3$ & $\pm    4$\\             

$e\sin \omega$       &$-0.05926$  & $-0.05763$ & $-0.06064$ & $-0.05806$\vspace{-0.8mm}\\  
                     & $\pm 176$  &  $\pm 186$ & $\pm  206$ & $\pm  256$\\                

$e$                  &  0.0761    &   0.0748   &   0.0772   &  0.0751\\                 
                                                         
$\omega$ \, (\degr)  &  309.88    &   309.65   &   308.21   & 309.39 \\                 
                            
$r_p$                &  0.0579    &   0.0579   &   0.0578   &  0.0578\\                 

$r_s$                &  0.0542    &   0.0543   &   0.0543   &  0.0546\\                 
                            
$k$                  &  0.9356    &   0.9391   &   0.9386   &   0.9446\vspace{-0.8mm}\\  
                     &  $\pm33$   &   $\pm43$  &   $\pm45$  &   $\pm59$\\                

$r_p + r_s$          &  0.1121   &   0.1122  &   0.1121  &  0.1123\vspace{-0.8mm}\\  
                     &  $\pm 2$  &   $\pm 2$ &   $\pm 2$ &  $\pm 3$ \\               
                            
$u_p$                &  0.56      &   0.66     &   0.74     &  0.76\\                   

$u_s$                &  0.59      &   0.68     &   0.77     &  0.81\\                   

$y_p$                &  0.37      &   0.42     &   0.48     &  0.57\\

$y_s$                &  0.38      &   0.43     &   0.50     &  0.59\\

$J_s/J_p$            &  0.8737    &    0.8415  &   0.8154   &  0.8037\vspace{-0.8mm}\\
                     &  $\pm25$   &    $\pm30$ &   $\pm33$  &  $\pm41$\\

$L_s/L_p$            &  0.7554    &    0.7357  &   0.7088   &  0.7011 \\

$\sigma$ \, (mmag.)  &  6.9       &    6.7     &   7.3      &  8.7   \\
\noalign{\smallskip}            
\hline
\end{tabular}            
\end{center}            
\textsc{Note 1:}
Linear limb darkening coefficients by Van Hamme (\cite{vh93}) were adopted,
a mass ratio of 0.97 was assumed, and phase shift and  magnitude normalization 
were included as free parameters

\textsc{Note 2:}
The errors quoted for the free parameters are the $formal$ errors determined 
from the iterative least squares solution procedure
\end{table}

Since \EW\ is well detached, the photometric elements have been determined 
from JKTEBOP analyses (Southworth et al. \cite{sms04a},
\cite{sms04b}) of the  $uvby$ light curves.
The underlying Nelson-Davis-Etzel binary model (Nelson \& Davis \cite{nd72},
Etzel \cite{e81}, Martynov \cite{m73}) represents the deformed stars
as biaxial ellipsoids and applies a simple bolometric reflection model.
We refer to Clausen et al. (\cite{avw08}, hereafter CTB08) 
for details on the binary model and code, and on the general approach applied.
In tables and text, we use the following symbols:
$i$ orbital inclination;
$e$ eccentricity of orbit;
$\omega$ longitude of periastron;
$r$ relative radius (in units of the semi-major axis);
$k = r_s/r_p$;
$u$ linear limb darkening coefficient;
$y$ gravity darkening coefficient;
$J$ central surface brightness;
$L$ luminosity;
$T_{\rm eff}$ effective temperature.

The mass ratio between the components was kept at the spectroscopic value;
see Sect.~\ref{sec:spel}.
The simple  built-in bolometric reflection model was used,
linear limb darkening coefficients by Van Hamme (\cite{vh93}) and  
Claret (\cite{c00}) were applied, or included as free parameters, and
gravity darkening coefficients corresponding to convective atmospheres
were adopted. 

Solutions for \EW\ are presented in Table~\ref{tab:ewori_jktebop_vh},
and as seen, the results from the four bands agree well.
Changing from Van Hamme to Claret limb darkening coefficients,
which are 0.06--0.10 higher, does not change the orbital and
stellar parameters significantly.
Coefficients determined from the light curves reproduce those by Van 
Hamme slightly better than those by Claret; they have formal uncertainties 
of about $\pm 0.05$. 
Including non-linear limb darkening (logarithmic or
square-root law) also has no significant effect on the photometric elements.

The adopted photometric elements listed in Table~\ref{tab:ewori_phel}
are the weighted mean values of the JKTEBOP solutions adopting the linear
limb darkening coefficients by Van Hamme. 
Realistic errors, based on 1\,000 Monte Carlo simulations in each band 
and on comparison between the $uvby$ solutions, have been assigned.
The Monte Carlo simulations include random variations within $\pm 0.07$
of the linear limb darkening coefficients.
As seen, the relative radii have been established to better than 0.5\%.
Due to the accurate light curves with about 265 points within eclipses -- 
coupled to the fact that secondary eclipse is total -- 
we obtain a reliable photometric determination of $k$, and the corresponding
luminosity ratios are exactly identical to those derived 
directly from the depths of the total secondary eclipse.
Also, $e$ and $\omega$ are much better constrained than from the
radial velocity analyses; see Sect.~\ref{sec:spel}.
At phase 0.0, about 90\% of the $y$ light from the primary component
is eclipsed.

For comparison, Popper et al. (\cite{plft86}) obtained 
$r_p = 0.0561 \pm 0.005$ and $k = 0.955 \pm 0.030$, an orbital
inclination of $i = 89\fdg65 \pm 0\fdg1$, and a $V$ luminosity
ratio of $L_s/L_p = 0.745 \pm 0.037$. 
Inside eclipses, only 69 $V,R$ points from two nights were
available, and to constrain $k$ they adopted a fixed 
$e\sin\omega = -0.0486$, based on $e = 0.068 \pm 0.004$ 
from the spectroscopic orbit and $e\cos\omega$ determined 
separately from the light curve analysis.

In conclusion, the new photometric elements differ somewhat from
those by Popper et al. and are significantly more accurate.

\begin{table}            
\caption[]{\label{tab:ewori_phel}
Adopted photometric elements for EW\,Ori.
}
\begin{center}             
\begin{tabular}{ll}             
\noalign{\smallskip}             
\hline             
\noalign{\smallskip}             
$i$              & $89{\fdg}86 \pm 0{\fdg}09$ \\ 
$e$              & $0.0758 \pm 0.0020$ \\       
$\omega$         & $309{\fdg}0 \pm  1{\fdg}3$ \\
$r_p$            & $0.0578 \pm 0.0002$ \\      
$r_s$            & $0.0543 \pm 0.0002$ \\     
$r_p + r_s$      & $0.1122 \pm 0.0002$ \\    
$k$              & $0.939\pm 0.005$ \\      
\noalign{\smallskip}             
\end{tabular}             
\begin{tabular}{lrrrr}             
\noalign{\smallskip}             
                 & $y$    & $b$    & $v$   & $u$  \\           
\noalign{\smallskip}             
$J_s/J_p$        & 0.873  & 0.843  & 0.814 & 0.805 \\
                 & $\pm24$&$\pm23$&$\pm22$ &$\pm23$\\
$L_s/L_p$        & 0.7601 & 0.7372 & 0.7087& 0.6943\vspace{-0.8mm}\\
                 & $\pm  53$&$\pm  59$&$\pm  56$ &$\pm  75$\\   
\noalign{\smallskip}             
\hline             
\end{tabular}             
\end{center}            
\textsc{Note:} The individual flux and luminosity ratios are based
on the mean stellar and orbital parameters
\end{table}

\begin{figure}
\epsfxsize=85mm
\epsfbox{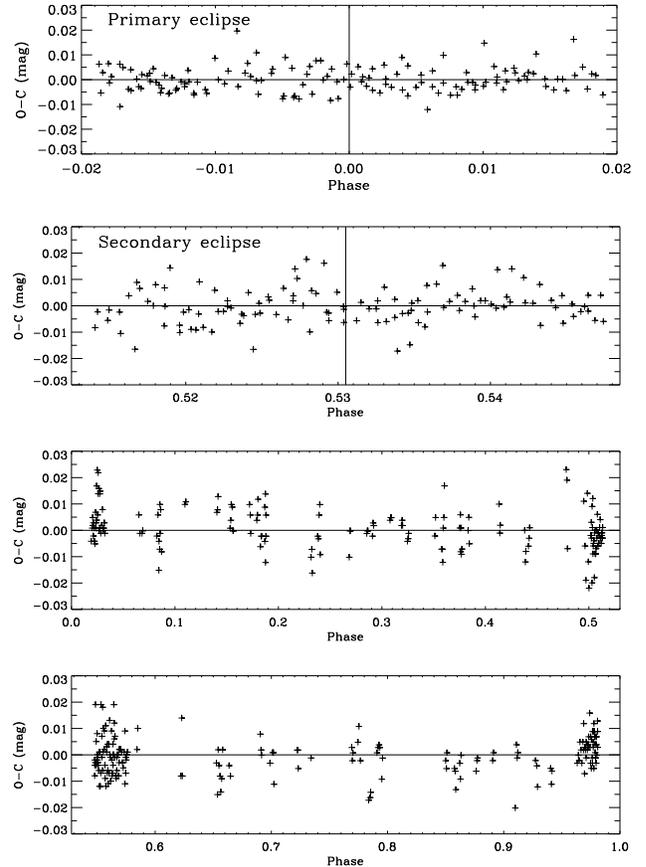}
\caption[]{\label{fig:ewori_res_y}
($O\!-\!C$) residuals of the \EW\ $y$-band observations from the theoretical
light curve computed for the photometric elements given in
Table~\ref{tab:ewori_jktebop_vh}.
}
\end{figure}

\section{Spectroscopic elements}
\label{sec:spel}

\begin{table}   
\caption[]{\label{tab:ewori_spel}
Spectroscopic orbital solutions for \EW\ determined from re-analyses of the
radial velocity observations by Popper et al. (\cite{plft86}) and Imbert (\cite{imbert02}).
}
\begin{center}    
\begin{tabular}{lrr} \hline   
\hline\noalign{\smallskip}    
Parameter:           & \multicolumn{1}{c}{Popper et al.}  & \multicolumn{1}{c}{Imbert (adopted)} \\ 
\noalign{\smallskip}
\hline
\noalign{\smallskip}    
Adjusted quantities:            &                  &                   \\ 
$K_p$~(\kms)                    &$ 73.18 \pm 0.41$ &  $72.26 \pm 0.24$  \\
$K_s$~(\kms)                    &$ 75.46 \pm 0.43$ &  $75.44 \pm 0.31$  \\
$\gamma_p$~(\kms)               &$ -7.03 \pm 0.37$ &  $-6.00 \pm 0.18$  \\
$\gamma_s$~(\kms)               &$ -7.21 \pm 0.39$ &  $-5.40 \pm 0.24$  \\
\noalign{\smallskip}  
Adopted quantities:             &                  & \\
$P$~(days)                      &  6.93684324      &  6.93684324 \\      
$T$~(HJD$-$2\,400\,000)$^{\mathrm{a}}$    &  51877.8014      & 51877.8014  \\ 
$T_p$~(HJD$-$2\,400\,000)$^{\mathrm{b}}$  &  51875.1968      & 51875.1968  \\
$e$                                       &  0.0758          & 0.0758      \\ 
$\omega$ \, (\degr)                       &  309.1           & 309.1       \\ 
\noalign{\smallskip}  
Derived quantities:               &                    &          \\
$M_p \sin^3i~\mathrm{(M_{\sun})}$ &$1.188 \pm 0.015$  &$1.173 \pm 0.010$ \\
$M_s \sin^3i~\mathrm{(M_{\sun}})$ &$1.152  \pm 0.014$ &$1.123 \pm 0.009$ \\
$q = M_s/M_p$                     &$0.970 \pm 0.008$  &$0.958 \pm 0.005$ \\
$a \sin i~\mathrm{(R_{\sun})}$    &$20.313 \pm 0.081$ &$20.184 \pm 0.054$\\
\noalign{\smallskip}  
Other quantities                         &      &      \\
pertaining to the fit:                   &      &      \\
$N_{obs} (p/s)$                          &17/17 &25/24 \\
Time span (days)                         &1897  &1857  \\
$\sigma_p$$^{\mathrm{c}}$ \,(\kms)       &  1.53& 0.91 \\               
$\sigma_s$$^{\mathrm{c}}$ \,(\kms)       &  1.62& 1.16 \\              
\noalign{\smallskip}  
\hline
\end{tabular}            
\begin{list}{}{}
\item[$^{\mathrm{a}}$] Time of periastron
\item[$^{\mathrm{b}}$] Time of central primary eclipse
\item[$^{\mathrm{c}}$] Standard deviation of a single radial velocity
\end{list}{}{}
\end{center}            
\end{table}

Spectroscopic orbits have been derived from re-analyses of the
radial velocities by Popper et al. (\cite{plft86}) and Imbert 
(\cite{imbert02}).
We have used the method of Lehman-Filh\'es implemented in the 
{\sc sbop}\footnote{Spectroscopic Binary Orbit Program, \\ 
{\scriptsize\tt http://mintaka.sdsu.edu/faculty/etzel/}}
program (Etzel \cite{sbop}), which is a modified and expanded version of
an earlier code by Wolfe, Horak \& Storer (\cite{wolfe67}).
The orbital period $P$ was fixed at the ephemeris value 
(Eq.~\ref{eq:ewori_eph}),
and the eccentricity $e$ and longitude of periastron $\omega$ at
the better constrained results from the photometric analysis 
(Table~\ref{tab:ewori_phel}).
Equal weights were assigned to the radial velocities of Popper et al.,
whereas the CORAVEL velocities by Imbert were weighted according to
the inverse square of their internal errors. 

\begin{figure}
\epsfxsize=095mm
\epsfbox{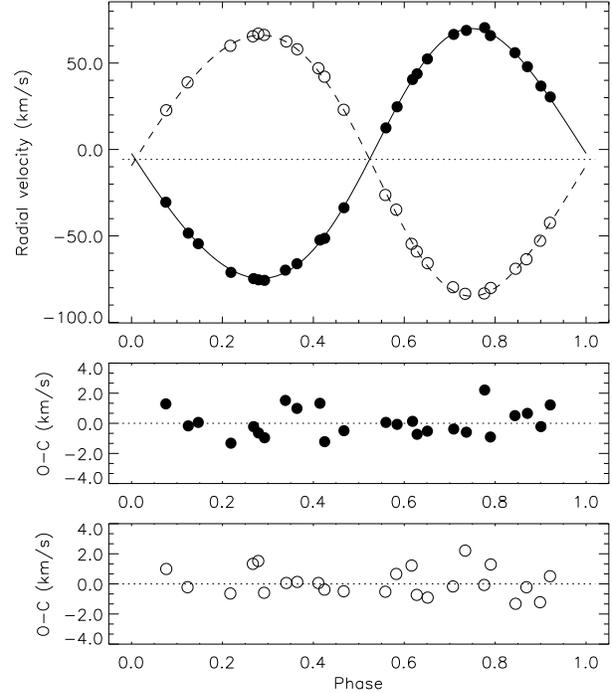}
\caption[]{\label{fig:ewori_sporb}
Adopted spectroscopic orbital solution for EW\,Ori (solid line: primary;
dashed line: secondary) and radial velocities (filled circles:
primary; open circles: secondary).
The dotted line (upper panel) represents the center-of-mass
velocity of the system.
Phase 0.0 corresponds to central primary eclipse.
}
\end{figure}

The spectroscopic elements are presented in Table~\ref{tab:ewori_spel}.
The radial velocities of the components were analysed independently
(SB1 solutions), but we note that SB2 solutions lead to nearly identical 
results.
Within errors, the semiamplitudes agree with the published results,
which are, however, based on different orbital periods and treatments of
$e$ and $\omega$. 
For the secondary component, the semiamplitudes of our two solutions
agree well, whereas they differ by almost 1 \kms\ for the primary.
We adopt the elements determined from the CORAVEL velocities, which
are more accurate and have a better phase coverage; 
see Fig.~\ref{fig:ewori_sporb}.

\section{Chemical abundances}
\label{sec:abund}

\begin{table}
\caption[]{\label{tab:feros}
Log of the FEROS observations of \EW.
}
\begin{minipage}{\columnwidth}
\centering
\renewcommand{\footnoterule}{}  
\begin{tabular}{ccrrr}
\hline
\hline\noalign{\smallskip}
HJD$-$2\,400\,000\footnote{Refers to mid-exposure}  & phase &t$_{exp}$\footnote{Exposure time in seconds} & S/N\footnote{Signal-to-noise ratio measured around 6070 {\AA}}\\
\noalign{\smallskip}
\hline
\noalign{\smallskip}
51209.62804        &  0.67760& 1800 &  100   \\   
51212.62786        &  0.11005& 2400 &  130   \\   
\hline
\end{tabular}
\end{minipage}
\end{table}

\begin{table}
\caption[]{\label{tab:ewori_abund}
Abundances ($[\mathrm{El./H}]$) for the primary and secondary
components of \EW\ determined from the two FEROS spectra.
}
\begin{center}
\begin{tabular}{lrlrrlr} \hline
\hline\noalign{\smallskip}
             &  \multicolumn{3}{c}{Primary} & \multicolumn{3}{c}{Secondary} \\
Ion          &[El./H]&  rms& N$_t$/N$_l$&[El./H]& rms & N$_t$/N$_l$  \\
\noalign{\smallskip}
\hline
Si\ione\     &  0.08 & 0.06& 12/10 &  0.03 & 0.11& 11/9 \\  
Ca\ione\     &  0.06 & 0.08&  8/6  &       &     &      \\  
Sc\itwo\     &  0.09 & 0.05&  4/4  &       &     &      \\  
Ti\ione\     &$-0.02$& 0.05&  4/3  &$-0.01$& 0.08&  8/8 \\  
Cr\ione\     &  0.09 & 0.13&  5/5  &$-0.01$& 0.10&  3/3\\  %
Fe\ione\     &  0.05 & 0.10&122/91 &  0.02 & 0.13& 92/74   \\
Fe\itwo\     &  0.03 & 0.09& 10/8  &  0.12 & 0.11& 8/7\\
Ni\ione\     &  0.09 & 0.10& 30/24 &  0.01 & 0.15& 19/14\\    %
\noalign{\smallskip}
\hline
\end{tabular}            
\end{center}
\textsc{Note:}
N$_t$ is the total number of lines used per ion, and N$_l$ is the
number of different lines used per ion.
Ions with at least 3 lines measured are included.
\end{table}

To determine the chemical composition of \EW, we have obtained 
two high-resolution spectra. They were observed at different phases 
and opposite line shifts with the FEROS fiber echelle spectrograph at ESO,
La Silla January--February 1999; see Table~\ref{tab:feros}. 
Details on the spectrograph, the reduction of the spectra,
and the basic approach followed in the abundance analyses is
described by CTB08.

The versatile VWA tool, now extended to analyses of double-lined
spectra, was used. We refer to Bruntt et al. 
(\cite{bruntt04,bruntt08}) and Bruntt (\cite{bruntt09}) 
for detailed descriptions of VWA.  It uses the SYNTH
software (Valenti \& Piskunov \cite{vp96}) to generate the synthetic spectra.
Atmosphere models were interpolated from the recent grid of MARCS
model atmospheres (Gustafsson et al. \cite{marcs08}), which adopt
the solar composition by Grevesse et al. (\cite{gas07}).
Line information was taken from the Vienna Atomic Line Database (VALD;
Kupka et al. \cite{kupka99}), but in order to derive abundances relative
to the Sun, log($gf$) values have been adjusted in such a way that each
measured line in the Wallace et al. (\cite{whl98}) solar atlas reproduces
the atmospheric abundances by Grevesse et al. (\cite{gas07}).
Analyses of a FEROS sky spectrum reproduce these adjustments closely.

The abundance results derived from all useful lines in both spectra
are presented in Table~\ref{tab:ewori_abund}.
We have only included lines with $measured$ equivalent
widths above 10 m{\AA} and below 50 m{\AA} (primary) and 40 m{\AA} (secondary).
The lines are diluted by a factor of about 1.8 (primary)
and 2.3 (secondary), meaning that lines with $intrinsic$ strengths
above 90 m{\AA} are excluded.
Comparing the results from the two spectra, we find no significant
differences. The effective temperatures, surface gravities
and rotational velocities listed in Table~\ref{tab:ewori_absdim} were adopted.
Microturbulence velocities were tuned until Fe \ione\ abundances were
independent of line equivalent widths, and the resulting values are 
$1.22 \pm 0.16$ (primary) and $1.45 \pm 0.30$ (secondary) \kms. 
The calibration by Edvardsson et al. (\cite{be93}) predicts 
$1.48 \pm 0.31$ \kms\ (primary) and $1.30 \pm 0.31$ (secondary).
For the adopted effective temperatures we see no dependency of the
abundance on excitation potential, which, however, occurs if they are changed
by more than about $\pm 150$~K.

As seen, a robust \feh\ is obtained, with identical
results from Fe\,\ione\ and Fe\,\itwo\ lines of both components. 
Changing the  model temperatures by $\pm 100$~K modifies \feh\
from the Fe\,\ione\ lines by about $\pm0.06$ dex whereas almost no
effect is seen for Fe\,\itwo\ lines.
If 0.25 \kms\ higher microturbulence velocities are adopted,
\feh\ decreases by about 0.06 dex for both neutral and ionized lines.
Taking these contributions to the uncertainties into account,
we adopt \feh\,$=+0.05\pm0.09$ for \EW.
  
We also find relative abundances close to +0.05 dex
for the other ions listed in Table~\ref{tab:ewori_abund}, including the
$\alpha$-elements Si, Ca, and Ti. 

As a supplement to the spectroscopic abundance analyses, we have
derived metal abundances from the de-reddened $uvby$ indices for the
individual components (Table~\ref{tab:ewori_absdim}) and
the calibration by Holmberg et al. (\cite{holmberg07}).
The results are: \feh\,$= 0.03\pm0.12$ (primary) and 
\feh\,$=-0.05\pm0.12$ (secondary). Within errors they agree with 
those from the spectroscopic analysis;
the quoted \feh\ errors include the uncertainties of the 
photometric indices and the published spread of the calibration.

\section{Absolute dimensions}
\label{sec:absdim}

Absolute dimensions for \EW\ are presented in Table~\ref{tab:ewori_absdim}, 
as calculated from the photometric and spectroscopic elements given in 
Tables~\ref{tab:ewori_phel} and \ref{tab:ewori_spel}.
As seen, masses and radii accurate to 0.9\% and 0.5\%, respectively,
have been established for the binary components.
For the radii, this is a clear improvement compared to the 1--3\% different
results listed in the new review on masses and radii by Torres et al. 
(\cite{tag09}), which are based on the work by Popper et al. (\cite{plft86})
and Imbert (\cite{imbert02}).

The $V$ magnitudes and $uvby$ indices for the components, included in
Table~\ref{tab:ewori_absdim}, were calculated from the the combined 
magnitudes and indices of the systems outside eclipses 
(Table~\ref{tab:ewori_std}) and the luminosity ratios between their 
components (Table~\ref{tab:ewori_phel}). 
The $V$ magnitude and the $uvby$ 
indices obtained for the primary component agree very well with those measured 
during the total part of central secondary eclipse (Table~\ref{tab:ewori_std}).

The $E(b-y)$ interstellar reddening, also given in
Table~\ref{tab:ewori_absdim}, was determined from the calibration by 
Olsen (\cite{olsen88}), using the $uvby\beta$ standard photometry for the 
combined light outside eclipses. For comparison, Popper et al. 
(\cite{plft86}) estimated $E(b-y) = 0.010 \pm 0.009$,
equivalent to the reddening listed by Torres et al. (\cite{tag09}),
$E(B-V) = 0.014 \pm 0.012$.
The model by Hakkila et al. (\cite{hak97}) yields a negative reddening in
the direction of and at the distance of \EW, whereas the maps by
Burstein \& Heiles (\cite{bh82}) and Schlegel et al. (\cite{sch98}) give
high {\it total} $E(B-V)$ reddenings of 0.09 and 0.16, respectively.
Knude (private communication) finds that \EW\ is located in or in front
of a tiny cloud in the outskirts of the Orion OB Ia association.

From the individual indices and the calibration by Holmberg et al. 
(\cite{holmberg07}), we derive effective temperatures of
$6070 \pm  95$ and $5870 \pm  95$~K for the primary and secondary
components, respectively, assuming the final \feh\ abundance. 
The temperature uncertainties include those of the $uvby$ indices, 
$E(b-y)$, \feh, and the calibration itself.
Temperatures based on the calibrations by Alonso et al. (\cite{alonso96})
and Ram\'irez \& Mel\'endez (\cite{rm05}) are slightly lower but
agree within errors. 
2MASS photometry of the combined light at phase 0.58 and the $V-K_s$
calibration by Masana et al. (\cite{masana06}) gives an 'average' temperature
of 6100~K compared to about~6000 K obtained from the combined $uvby$ indices.
Finally, the empirical flux scale by Popper (\cite{dmp80}) 
and the $y$ flux ratio between the components (Table~\ref{tab:ewori_phel})
yield a well-established temperature difference between the components of
$170 \pm 30$~K (excluding possible errors of the scale itself).
Consequently, we assign temperatures of 6070 and 5900~K.
They are about 100~K higher than adopted by Popper et al. (\cite{plft86})
and Torres et al. (\cite{tag09}).

\begin{table}   
\caption[]{\label{tab:ewori_absdim}
Astrophysical data for EW\,Ori.
}
\begin{minipage}{\columnwidth}
\begin{center}    
\begin{tabular}{lrr} \hline    
\noalign{\smallskip}    
\hline    
\noalign{\smallskip}    
                     &    Primary       &    Secondary      \\ 
\noalign{\smallskip}    
\hline    
\noalign{\smallskip}    
Absolute dimensions:          &                   &                 
 \\ 
$M/M_{\sun}$                  &$1.173 \pm 0.011$  &$1.123 \pm 0.009$
\\ 
$R/R_{\sun}$                  &$1.168 \pm 0.005$  &$1.097 \pm 0.005$ 
\\ 
$\log g$ (cgs)                & $4.372 \pm 0.005$ & $4.408 \pm 0.005$
\\
$v \sin i$$^{\mathrm{a}}$ (\kms)  & $9.0\pm0.7$       & $8.8\pm0.6$
\\ 
$v_{sync}$$^{\mathrm{b}}$ (\kms)  & $ 8.5 \pm 0.0$    & $ 8.0 \pm 0.0$
\\ 
$v_{psync}$$^{\mathrm{c}}$ (\kms) & $ 8.8 \pm 0.0$    & $ 8.3 \pm 0.0$
\\ 
$v_{peri}$$^{\mathrm{d}}$ (\kms)  & $ 9.9 \pm 0.1$    & $ 9.3 \pm 0.1$
\\ 
 & & \\
Photometric data:             &                   &                 
\\ 
$V$$^{\mathrm{e}}$ &         $10.516 \pm 0.008$  &        $10.814 \pm 0.008$\\
$(b-y)$$^{\mathrm{e}}$ &     $ 0.376 \pm 0.006$  &        $ 0.409 \pm 0.006$\\
$m_1$$^{\mathrm{e}}$   &     $ 0.182 \pm 0.010$  &        $ 0.192 \pm 0.010$\\
$c_1$$^{\mathrm{e}}$   &     $ 0.373 \pm 0.011$  &        $ 0.352 \pm 0.012$\\
$E(b-y)$    & \multicolumn{2}{c}   {$0.019 \pm 0.010$} \\ 
$T_{\mbox{\scriptsize eff}}\,$      &  $6070 \pm  95$    &   $5900 \pm  95$ \\ 
$M_{\mbox{\scriptsize bol}}\,$      &  $4.19  \pm 0.07$  &   $4.45  \pm 0.07$ \\
$\log L/L_{\sun}$ & $0.22 \pm 0.03$ &    $0.12 \pm 0.03$ \\   
$BC$              & $-0.04$         &    $-0.06$ \\    
$M_V$ &             $ 4.23 \pm 0.07$&   $ 4.51 \pm 0.07$ \\ 
$V_0-M_V$         &$ 6.21  \pm 0.08 $& $ 6.22  \pm 0.09 $ \\
Distance \, (pc) &$ 175   \pm   7  $& $ 176   \pm   7  $ \\   
 & & \\ 
Abundance:                    &                   &   \\              
\feh\       & \multicolumn{2}{c}   {$+0.05 \pm 0.09$} \\
\noalign{\smallskip}            
\hline
\noalign{\smallskip}
\end{tabular}            
\begin{list}{}{}
\item[$^{\mathrm{a}}$] Observed rotational velocity
\item[$^{\mathrm{b}}$] Equatorial velocity for synchronous rotation
\item[$^{\mathrm{c}}$] Equatorial velocity for pseudo-synchronous rotation
\item[$^{\mathrm{d}}$] Refers to periastron velocity
\item[$^{\mathrm{e}}$] Not corrected for interstellar absorption/reddening
\end{list}
\end{center}
\textsc{Note:}
Bolometric corrections ($BC$) by Flower (\cite{flower96}) have been
assumed, together with
$T_{eff\sun} = 5780$ K, $BC_{\sun} = -0.08$, and $M_{bol\sun} = 4.74$.
\end{minipage}
\end{table}

The projected rotational velocities listed in Table~\ref{tab:ewori_absdim} were
determined from broadening function analyses 
(e.g. Kaluzny et al. \cite{k06}) of several orders of the
two FEROS spectra (Table~\ref{tab:feros}). Within errors they
agree with the synchronous and the pseudo-synchronous values (Hut \cite{hut81}, Eq. 42).
The turbulent dissipation and radiative damping formalism of Zahn
(\cite{zahn77,zahn89}) predicts
synchronization times scales of $3.4 \times 10^8$ yr (primary) and
$3.6 \times 10^8$ yr (secondary),
and a time scale for circularization of $7.4 \times 10^9$ yr,
compared to predicted age of \EW\ of about $2 \times 10^9$ yr 
(Sect.~\ref{sec:dis}).

The distance to \EW\ was calculated from the "classical" relation
(see e.g. CTB08), adopting the solar values and bolometric corrections given in
Table~\ref{tab:ewori_absdim}, and $A_V/E(b-y) = 4.28$ 
(Crawford \& Mandwewala \cite{cw76}).
As seen, identical values are obtained for the two components, and 
the distance has been established to 4\%, accounting for all
error sources and including the use of other $BC$ scales
(e.g. Code et al. \cite{code76}, Bessell et al. \cite{bessell98},
Girardi et al. \cite{girardi02}).
The empirical $K$ surface brightness - $T_{\rm eff}$ relation
by Kervella et al. (\cite{kervella04}) leads to nearly identical and perhaps
even more precise distances (about $\pm 3$~pc); 
see Southworth et al. (\cite{southworth05}) for details.

As mentioned in Sect.~\ref{sec:ewori}, \EW\ has been considered as a
possible member of Collinder 70. According to Kharchenko el al.
(\cite{kh05}) the distance to this open cluster is 391~pc, its radial 
velocity is +19.49 \kms, and its age is 5.1 Gyr. 
This rules out that \EW\ is a member; see Tables~\ref{tab:ewori_absdim}, 
\ref{tab:ewori_spel}, and Sect.~\ref{sec:dis}.

\section{Stellar activity}
\label{sec:act}

Popper et al. (\cite{plft86}) reported weak evidence for intrinsic variability
in their $V,R$ light curves. We see no clear signs of periodic and/or yearly
variations, e.g. due to spots, in the $uvby$ light curves, but
on the other hand the rms of the $O-C$ residuals of the $uvby$ observations 
from the theoretical light curves (Table~\ref{tab:ewori_jktebop_vh}) are 
marginally higher than expected from the average accuracy per point 
(Sect.~\ref{sec:lc}).
The Rossby\footnote{Defined as the ratio of the rotation
period to the convective turnover time.} numbers for the primary and
secondary components are approximately 0.90 and 0.42, respectively. This
places the secondary in the range where other stars tend to show spot
activity (Hall \cite{hall94}). 
Also, the FEROS spectra (see Sect.~\ref{sec:abund}) reveal weak 
Ca\,\itwo\ H and K emission from the secondary component\footnote{
Popper et al. (\cite{plft86}) saw no emission, probably due to inadequate
resolution.}. 
The secondary component is therefore likely to be active, but 
at a rather low level. There is no
information on X-ray emission from ROSAT (Voges et al. \cite{rosat99}).

\section{Discussion}
\label{sec:dis}

In the following, we compare the absolute dimensions obtained
for \EW\ with properties of recent theoretical stellar evolutionary models.
A detailed comparison with other similar, well-studied eclipsing binaries
will be included in a forthcoming paper.

\begin{figure}
\epsfxsize=90mm
\epsfbox{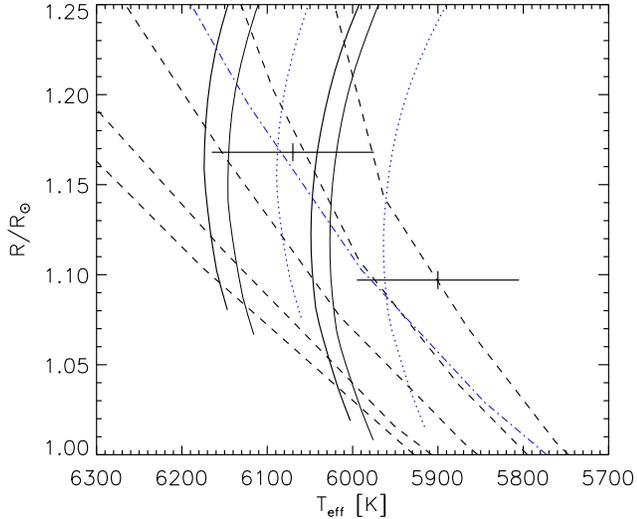}
\caption[]{\label{fig:ewori_tr}
\EW\ compared with $Y^2$ models for \feh\,$=+0.05$.
Tracks for the component masses (solid lines, thick) and isochrones for 0.5 and
1.0--4.0 Gyr (dashed; step 1.0 Gyr) are shown.
The uncertainty in the location of the tracks coming from
the mass errors are indicated (solid lines, thin).
Tracks (dotted, blue) and the 2 Gyr isochrone (dash-dot, thin, blue) 
for \feh\,$=+0.14$ are included.
}
\end{figure}

\begin{figure}
\epsfxsize=90mm
\epsfbox{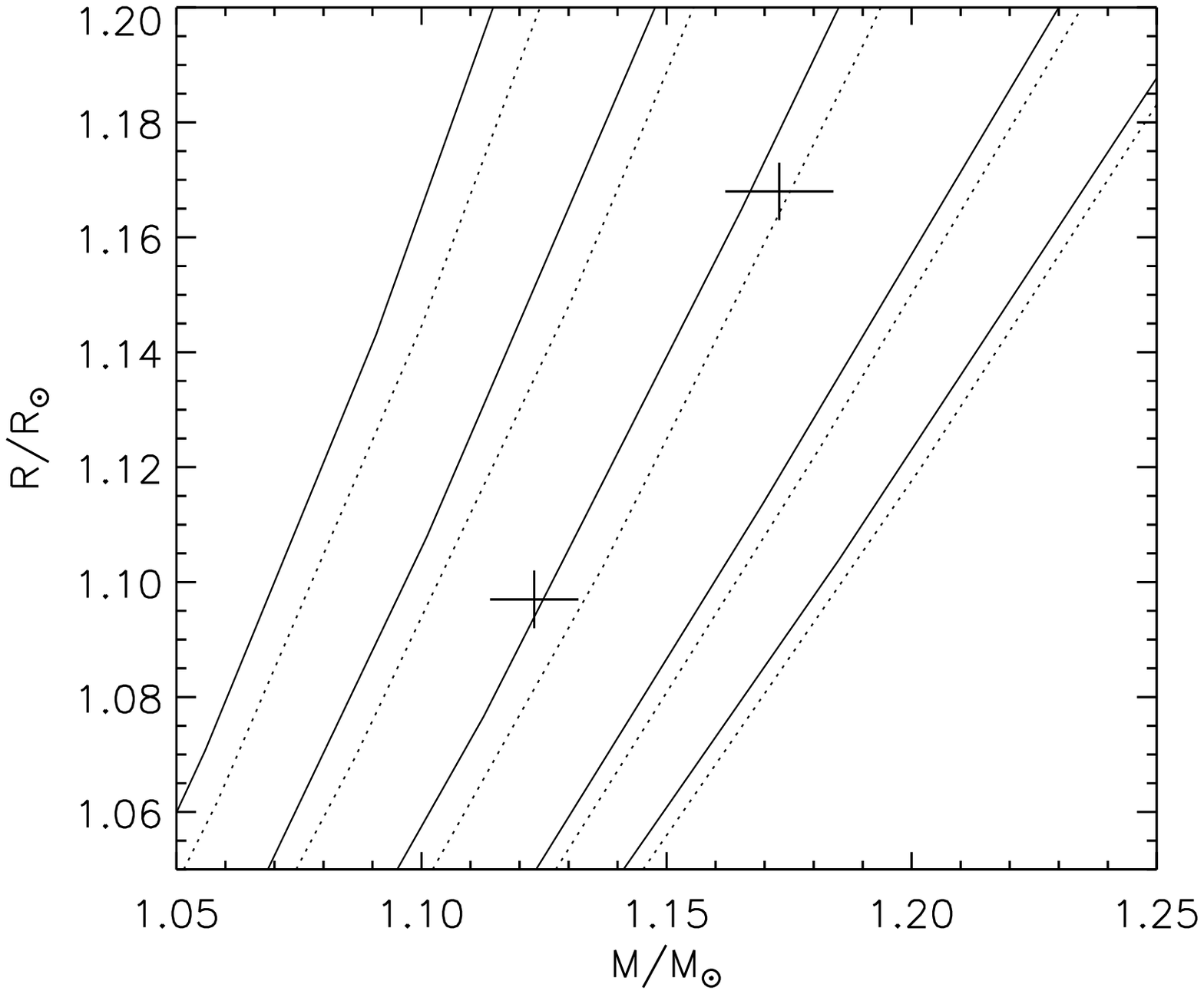}
\caption[]{\label{fig:ewori_mr}
\EW\ compared with $Y^2$ models for 
\feh\,$=+0.05$ (solid lines) and \feh\,$=+0.14$ (dotted).
Isochrones for 0.5 and 1.0--4.0 Gyr (step 1.0 Gyr) are shown.
}
\end{figure}

\begin{figure}
\epsfxsize=90mm
\epsfbox{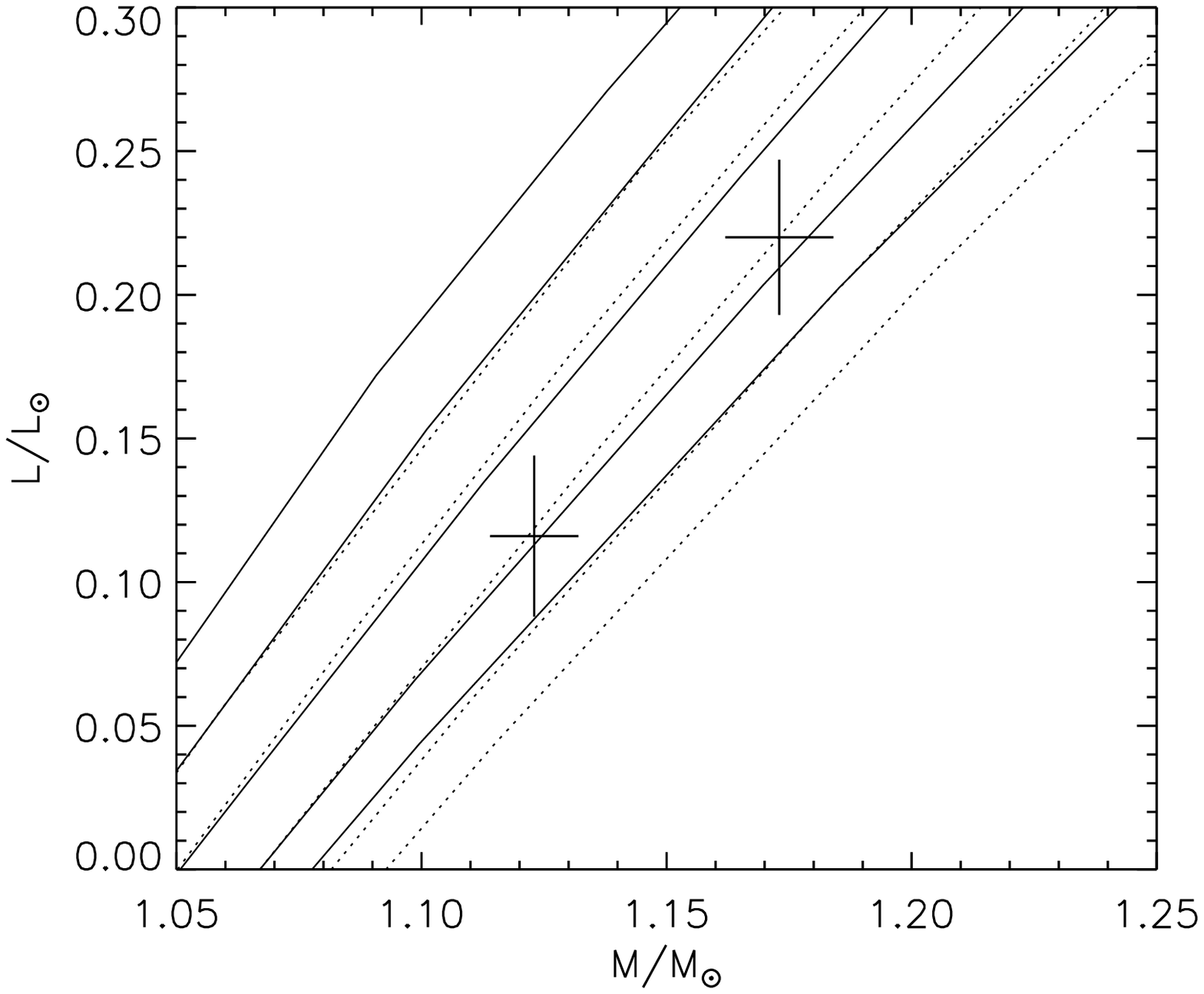}
\caption[]{\label{fig:ewori_ml}
\EW\ compared with $Y^2$ models for 
\feh\,$=+0.05$ (solid lines) and \feh\,$=+0.14$ (dotted).
Isochrones for 0.5 and 1.0--4.0 Gyr (step 1.0 Gyr) are shown.
}
\end{figure}

Figs.~\ref{fig:ewori_tr}, \ref{fig:ewori_mr},
and \ref{fig:ewori_ml} illustrate the results from comparisons with the
Yonsei-Yale ($Y^2$) evolutionary tracks and isochrones by Demarque et al.
(\cite{yale04})\footnote{{\scriptsize\tt http://www.astro.yale.edu/demarque/yystar.html}}.
The mixing length parameter in convective envelopes is calibrated using
the Sun, and is held fixed at $l/H_p = 1.7432$.
The enrichment law $Y = 0.23 + 2Z$ is adopted, together with the solar
mixture by Grevesse et al. (\cite{gns96}), leading to
($X$,$Y$,$Z$)$_{\sun}$ = (0.71564,0.26624,0.01812).
Only models for \afe\,$=0.0$ have been considered.
We refer to CTB08 for a brief description of other aspects of their
up-to-date input physics.

\begin{table*}
\caption[]{\label{tab:ewori_ac}
Information on the Claret models and ages inferred from the observed radii;
see Figs.~\ref{fig:ewori_tr_ac} and \ref{fig:ewori_ar_ac}.
}
\centering
\begin{tabular}{lcccccc} \hline
\hline\noalign{\smallskip}
Model/      &  $Y$  &  $Z$  &  $l/H_p$ & $l/H_p$    & Age (Gyr) & Age (Gyr)\\
Linestyle   &       &       &  Primary & Secondary  & Primary   & Secondary\\
\noalign{\smallskip}
\hline
\noalign{\smallskip}
1 dashed (red)        & 0.280 & 0.020&  1.68 & 1.68   & $1.90 \pm 0.10$ & $2.16 \pm 0.10$\\
2 full, thin (blue)   & 0.280 & 0.020&  1.54 & 1.50   & $1.45 \pm 0.10$ & $1.45 \pm 0.10$ \\
3 full, thick (black) & 0.270 & 0.020&  1.68 & 1.60   & $2.28 \pm 0.10$ & $2.28 \pm 0.10$\\
\hline
\end{tabular}
\end{table*}

As seen from Fig.~\ref{fig:ewori_tr}, models for the observed
masses and abundance, \feh\,$=+0.05$, equivalent to
($X$,$Y$,$Z$) = (0.70955,0.27030,0.02015),
are hotter than observed. 
Also, the well-established temperature difference between 
the components,$170 \pm 30$ K, is slightly larger than between 
the corresponding models, although this is partly covered by the uncertainty 
in the track positions coming from the small 0.9\% mass errors.
The uncertainty of \feh\ is 
$\pm 0.09$ dex, and models for \feh\,$=+0.14$, equivalent to 
($X$,$Y$,$Z$) = (0.69695,0.27870,0.02435), fit the components 
better. 
However, for \EW, the correlation between \feh\ and $T_{\rm eff}$ is
such, that a 0.09 dex higher metal abundance corresponds to 
150~K higher temperatures (see Sect.~\ref{sec:abund}), meaning that 
\feh\,$=+0.14$ models actually tend to become a bit too cool.
The best match is obtained for the combination of 75~K higher 
temperatures, e.g. coming from a 0.01 mag higher interstellar
reddening, and a 0.04 dex higher \feh. 

Turning to the scale-independent masses and radii, Fig.~\ref{fig:ewori_mr}
shows that the models predict nearly identical ages, close to 2 Gyr for
the components, perhaps with a slight tendency of a higher value for the 
secondary.
From the observed masses and luminosities, the models predict identical
but less precise ages of about 1 Gyr (\feh\,$=+0.05$) and 2 Gyr 
(\feh\,$=+0.14$); see Fig.~\ref{fig:ewori_ml}. 
Comparisons with solar-scaled (VRSS) Victoria-Regina models 
(VandenBerg et al., \cite{vr06}) lead to nearly identical results.     

\begin{figure}
\epsfxsize=90mm
\epsfbox{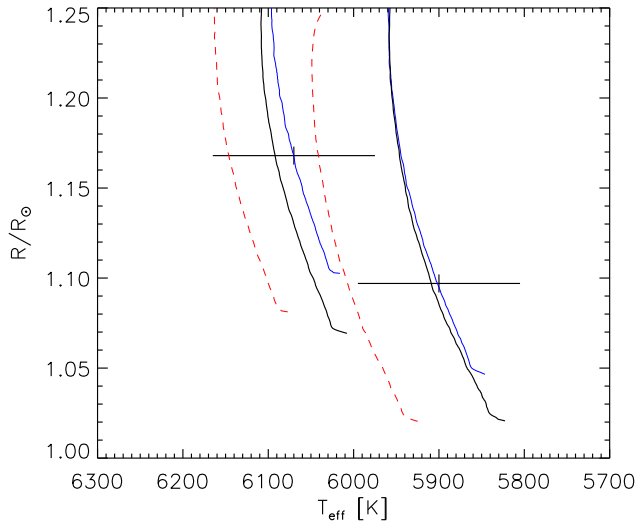}
\caption[]{\label{fig:ewori_tr_ac}
\EW\ compared to Claret models for the observed
masses and \feh\ abundance. See Table~\ref{tab:ewori_ac}
for details and linestyles/colours.
}
\end{figure}

\begin{figure}
\epsfxsize=90mm
\epsfbox{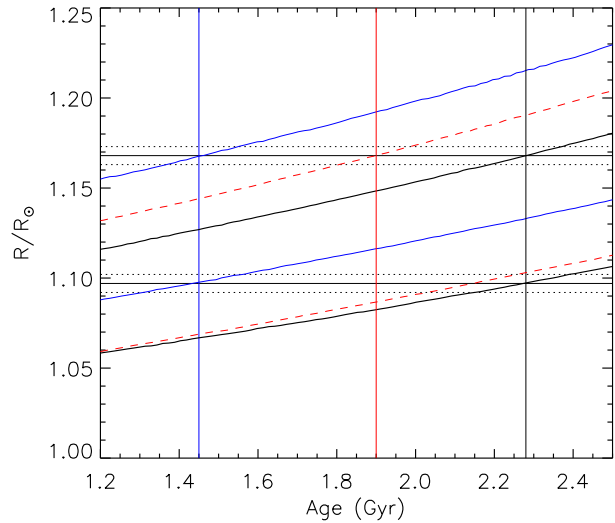}
\caption[]{\label{fig:ewori_ar_ac}
\EW\ compared to Claret models for the observed
masses and \feh\ abundance. See Table~\ref{tab:ewori_ac}
for details.
The curves illustrate model radii as function of age for the
components (upper: primary; lower: secondary).
The horizontal full drawn lines mark the observed radii of the
components with errors (dotted lines).
The vertical lines mark the ages predicted for the primary  
component.
}
\end{figure}

In conclusion, solar-scaled models provide fairly acceptable fits to 
the observed properties of \EW\ within their uncertainties. It is, however,
of interest to see if specific model tunings can lead to perfect reproduction
of \EW.  To that end, we have calculated
dedicated models for the observed \feh\ using the Granada code by Claret 
(\cite{c04}), which assumes an enrichment law of $Y = 0.24 + 2.0Z$ 
together with the solar mixture by Grevesse \& Sauval (\cite{gs98}).
The observed \feh\,$=+0.05$ then corresponds to ($X$,$Y$,$Z$) = 
(0.70,0.28,0.02).
The envelope mixing length parameter needed to reproduce the Sun
is $l/H_p = 1.68$, and the adopted amount of core overshooting
is $\alpha_{ov}$ = 0.20 (in units of the pressure 
scale height)\footnote{Models without core overshooting are very similar at
the age of \EW.}.

Table~\ref{tab:ewori_ac} lists the few models we have investigated, and
they are compared to \EW\ in Figs.~\ref{fig:ewori_tr_ac} and
\ref{fig:ewori_ar_ac}.
The model set \#1, which is closest to the $Y^2$ models for \feh\,$=+0.05$,
show the same temperature and age differences as discussed above, 
although the track shapes are somewhat different.  
By decreasing $l/H_p$ by 0.14 (primary) and 0.18 (secondary), the
models (set \#2) match \EW\ perfectly at an age of 1.45 Gyr.
Alternatively, keeping the solar $l/H_p$ for the primary and
decreasing it slightly by 0.08 for the secondary (model set \#3) also gives
a good fit, provided the helium content is lowered to $Y \approx0.27$,
i.e. close to the $Y^2$ value. The predicted age is 2.3 Gyr.
For the secondary, which exhibits signs of activity at a mild level
(see Sect. ~\ref{sec:lc}),
a lower $l/H_p$ -- and thereby larger model radius and lower model 
temperature -- is consistent with findings for other active solar-type
binary components (CBC09). On the other hand, for the primary there is no 
observational background for a lower $l/H_p$, and besides, 
2D radiation hydrodynamics calculations (Ludwig et al. \cite{hgl99}) 
predict mixing length parameters close to the solar value for inactive 
stars with temperatures and surface gravities like those of the
\EW\ components. So, the model set \#3 is our preferred fit.

Indications of a need for a slight 
downwards revision of the helium content, compared to the $Y-Z$ relations
adopted for the model grids, has been seen in a few other cases (e.g.
\object{VZ~Hya}, CTB08; \object{V1130~Tau}, in prep.). It is, however, still 
too early to drawn any firm conclusions. We will return to this issue
in forthcoming papers on analyses of several new solar-type binaries;
see the list in CBC09.

\section{Summary and conclusions}
\label{sec:sum}

From state-of-the-art observations and analyses, precise (0.5--0.9\%) absolute
dimensions have been established for the components of the totally eclipsing
G0~V system \EW.
A detailed spectroscopic analysis yields an iron abundance relative to the Sun
of \feh\,$=+0.05\pm0.09$ and similar relative abundances for Si, Ca, Sc, Cr,
and Ni.

The 1.12 \Msun\ secondary component reveals weak Ca\,\itwo\ H and K emission 
and is probably mildly active; we see no signs of activity for the 1.17 
\Msun\ primary.
Apsidal motion ($U = 16\,300 \pm 3\,900$ yr) with a 62\% relativistic 
contribution has been detected for the eccentric orbit 
($e = 0.0758 \pm 0.0020$), and the inferred mean central density concentration 
coefficient, log($k_2$) = $-1.66 \pm 0.30$, agrees marginally with model 
predictions.
The measured rotational velocities, $9.0 \pm 0.7$ (primary) and
$8.8 \pm 0.6$ (secondary) \kms, are in agreement with both synchronous rotation
and the theoretically predicted pseudo-synchronous velocities. 

Stellar models with solar-scaled envelope mixing length parameters
reproduce the observed properties of \EW\ fairly well at an age of 
$\approx$2 Gyr.
We demonstrate, however, that perfect agreement can be obtained by
$a)$ a slight downwards adjustment of the envelope mixing length parameter
for the secondary, as seen for other active solar-type stars, and
$b)$ a slightly lower helium content than prescribed by the $Y-Z$ relations
adopted for the various standard model grids.

This study is part of a larger project on solar-type eclipsing binaries;
see e.g. CBC09.

\begin{acknowledgements}
It is a great pleasure to thank the many colleagues and students,
who have shown interest in our project and have participated in the
extensive (semi)automatic observations of \EW\ at the SAT:
Gwillerm Berard,
Vanessa Doublier,
Mathias P. Egholm,
Anders Johansen,
Erling Johnsen,
Helene J{\o}rgensen,
Raslan Leguet,
Gilbert Mahoux, and
John D. Pritchard. 
Excellent technical support was received from the staffs of Copenhagen
University and ESO, La Silla.
We thank J.~M. Kreiner for providing a complete list of published times 
of eclipses for \EW\ and J. Southworth for access to his JKTEBOP code.
G. Torres and J. Knude kindly made independent interstellar reddening
information available.

The projects "Stellar structure and evolution -- new challenges from
ground and space observations" and "Stars: Central engines of the evolution
of the Universe", carried out at Copenhagen University and Aarhus University,
are supported by the Danish National Science Research Council.

The following internet-based resources were used in research for
this paper: the NASA Astrophysics Data System; the SIMBAD database
and the VizieR service operated by CDS, Strasbourg, France; the
ar$\chi$iv scientific paper preprint service operated by Cornell University;
the VALD database made available through the Institute of Astronomy,
Vienna, Austria; the MARCS stellar model atmosphere library.
This publication makes use of data products from the Two Micron
All Sky Survey, which is a joint project of the University of
Massachusetts and the Infrared Processing and Analysis Center/
California Institute of Technology, funded by the National
Aeronautics and Space Administration and the National Science
Foundation.
\end{acknowledgements}

{}

\listofobjects
\end{document}